\documentclass[preprint,pra,superscriptaddress]{revtex4-1}

\usepackage[dvips]{graphicx} % for figures
\usepackage{amsfonts}
\usepackage{amssymb}
\usepackage{amscd}
\usepackage{amsmath}    % need for subequations
\usepackage{enumerate}
\usepackage{epsfig}
\usepackage{subfigure}

\newcommand{\ket}[1]{\mbox{$\left| #1 \right\rangle$}}

\usepackage{amscd}
\usepackage{amsmath,amssymb,amsthm,mathrsfs,amsfonts,dsfont}
\usepackage{subfigure, epsfig}
\usepackage{braket}
\usepackage{bm}
\usepackage{enumerate}

\usepackage{graphicx}
\usepackage{epsfig}
\usepackage{epstopdf}
\usepackage{subfigure}
\usepackage{color}

\newcommand\oprod[2]{\ensuremath{|#1\rangle\langle#2|}}
\newcommand\mean[1]{\ensuremath{\langle #1 \rangle}}

\begin{document}

\title{Twin-Field Quantum Key Distribution over 511 km Optical Fiber Linking two Distant Metropolitans}

\author{Jiu-Peng Chen}
\email{The authors contributed equally to the paper.}
\author{Chi Zhang}
\email{The authors contributed equally to the paper.}
\author{Yang Liu}
\affiliation{Hefei National Laboratory for Physical Sciences at Microscale and Department of Modern Physics, University of Science and Technology of China, Hefei 230026, China}
\affiliation{Shanghai Branch, CAS Center for Excellence and Synergetic Innovation Center in Quantum Information and Quantum Physics, University of Science and Technology of China, Shanghai 201315, China}
\affiliation{Jinan Institute of Quantum Technology, Jinan, Shandong 250101, China}
\author{Cong Jiang}
\affiliation{Jinan Institute of Quantum Technology, Jinan, Shandong 250101, China}
\affiliation{State Key Laboratory of Low Dimensional Quantum Physics, Department of Physics, Tsinghua University, Beijing 100084, China}
\author{Wei-Jun Zhang}
\affiliation{State Key Laboratory of Functional Materials for Informatics, Shanghai Institute of Microsystem and Information Technology, Chinese Academy of Sciences, Shanghai 200050, China}
\author{Zhi-Yong Han}
\author{Shi-Zhao Ma}
\affiliation{Jinan Institute of Quantum Technology, Jinan, Shandong 250101, China}
\author{Xiao-Long Hu}
\affiliation{State Key Laboratory of Low Dimensional Quantum Physics, Department of Physics, Tsinghua University, Beijing 100084, China}
\author{Yu-Huai Li}
\author{Hui Liu}
\affiliation{Hefei National Laboratory for Physical Sciences at Microscale and Department of Modern Physics, University of Science and Technology of China, Hefei 230026, China}
\affiliation{Shanghai Branch, CAS Center for Excellence and Synergetic Innovation Center in Quantum Information and Quantum Physics, University of Science and Technology of China, Shanghai 201315, China}
\author{Fei Zhou}
\affiliation{Jinan Institute of Quantum Technology, Jinan, Shandong 250101, China}
\author{Hai-Feng Jiang}
\author{Teng-Yun Chen}
\affiliation{Hefei National Laboratory for Physical Sciences at Microscale and Department of Modern Physics, University of Science and Technology of China, Hefei 230026, China}
\affiliation{Shanghai Branch, CAS Center for Excellence and Synergetic Innovation Center in Quantum Information and Quantum Physics, University of Science and Technology of China, Shanghai 201315, China}
\author{Hao Li}
\author{Li-Xing You}
\author{Zhen Wang}
\affiliation{State Key Laboratory of Functional Materials for Informatics, Shanghai Institute of Microsystem and Information Technology, Chinese Academy of Sciences, Shanghai 200050, China}
\author{Xiang-Bin Wang}
\affiliation{Shanghai Branch, CAS Center for Excellence and Synergetic Innovation Center in Quantum Information and Quantum Physics, University of Science and Technology of China, Shanghai 201315, China}
\affiliation{Jinan Institute of Quantum Technology, Jinan, Shandong 250101, China}
\affiliation{State Key Laboratory of Low Dimensional Quantum Physics, Department of Physics, Tsinghua University, Beijing 100084, China}
\author{Qiang Zhang}
\author{Jian-Wei Pan}
\affiliation{Hefei National Laboratory for Physical Sciences at Microscale and Department of Modern Physics, University of Science and Technology of China, Hefei 230026, China}
\affiliation{Shanghai Branch, CAS Center for Excellence and Synergetic Innovation Center in Quantum Information and Quantum Physics, University of Science and Technology of China, Shanghai 201315, China}

\begin{abstract}
The basic principle of quantum mechanics~\cite{1982Wootters} guarantee the unconditional security of quantum key distribution (QKD)~\cite{BB84,mayers1996quantum,lo1999unconditional,shor2000simple,Scarani2009} at the cost of inability of amplification of quantum state. As a result, despite remarkable progress in worldwide metropolitan QKD networks~\cite{Gisin2002,RevModPhys.92.025002} over the past decades, long haul fiber QKD network without trustful relay has not been achieved yet. Here, through sending-or-not-sending (SNS) protocol~\cite{wang2018sns}, we complete a twin field QKD (TF-QKD)~\cite{nature18Overcoming} and distribute secure keys without any trusted repeater over a 511 km long haul fiber trunk linking two distant metropolitans. Our secure key rate is around 3 orders of magnitudes greater than what is expected if the previous QKD field test system over the same length were applied. The efficient quantum-state transmission and stable single-photon interference over such a long distance deployed fiber paves the way to large-scale fiber quantum networks.
\end{abstract}

\maketitle

\section*{Introduction}
Quantum non-cloning theorem~\cite{1982Wootters} forbids perfect cloning of unknown quantum states or non-orthogonal states simultaneously. Based on the theorem, the first QKD protocol, BB84~\cite{BB84} has been proposed, in which non-orthogonal states are exploited to encode random key and distributed between two authorized users. Any probe from an unauthorized party intending to steal the key can be seen as a kind of clone, which will inevitably bring additional bit error according to the non-cloning theorem. This bit error can be found during the authorized users' post processing and thus the information theoretical security is achieved~\cite{mayers1996quantum,lo1999unconditional,shor2000simple}.

After that, QKD has been extensively studied~\cite{BB84,Gisin2002,Scarani2009,RevModPhys.92.025002} and grown to a matured technology in a real-world application today. Quite a few metropolitan fiber network has been built all over the world~\cite{peev2009the,stucki2011longterm,chen2009field,dynes2019cambridge,chen2010metropolitan,wang2010field}. Similar as the optical fiber communication (OFC), the quantum state will also be attenuated exponentially with the transmission distance. For OFC, optical amplifier is used to rely the optical signal every 80 km~\cite{ofc2017} to build up a long haul fiber network. The amplifier for single unknown quantum state, however doesn't exist because the amplification of quantum signal can also be seen as a type of clone and any cloning of unknown quantum states will bring fatal errors~\cite{1982Wootters}. Therefore, a simple optical amplifier can't be used for long haul fiber QKD. So far the longest reported field test of QKD is around 90~km~\cite{natureyuao}.

Trustful relay is utilized to build up the long haul trunk QKD~\cite{natureyuao} but the many relay stations must be well isolated and trustful. Meanwhile, quantum repeater~\cite{dlcz,yuentanglement2020} has been invented to replace optical amplifier. But the current technology can only reach 50 km~\cite{weinfurter2020long,liquantum2020}. Recently, it has been shown that there exists an upper bound for both repeater-less QKD or OFC~\cite{PLOB2017}. Therefore, the current main challenge towards large scale fiber quantum communication network is to beat the repeater-less bound and demonstrate QKD over real long haul trunk line.

Among all protocols, TF QKD~\cite{nature18Overcoming} promises high key rates over long distances to beat the repeater-less bound, which enhances the key rate to a square-root scaling ($\sqrt{\eta}$) to the channel transmittance. Significantly, the experimental demonstrations of TF-QKD~\cite{minder2019experimental,liu2019experimental,wang2019beating,zhong2019proof,fang2019surpassing,chen2020sending} achieved a record-long distribution distance of more than 500 km~\cite{fang2019surpassing,chen2020sending}. However, all these experiments were implemented in a laboratory, leaving the question that whether this fancy protocol is feasible in a practical scenario.

In general, comparing to lab experiments, field tests in OFC network often introduce more noise due to complex environmental fluctuation and crosstalk from adjacent classical fiber communication. The condition becomes even more challenging for TF-QKD, which requires optical phase stability for single photon level interference. For example, in the lab, a vibration due to human voice or walk will influence the optical phase and decrease the interference visibility~\cite{chen2020sending}. While, in the field, the human voice or activity can't be forbidden like in the lab. Considering these challenges, putting the experiment to a field test is not only necessary to prove its feasibility, but also an important step to explore the possibilities of the future global QKD networks.

Here, we have precisely controlled the wavelength of two 500 km away independent laser sources and fast compensated any small phase fluctuation in the channel. Then, we present a field test of a TF-QKD experiment through a total channel length of 511 km ultra-low-loss fiber (including 430 km long haul fiber and 81 km fiber spool) with a total loss of 89.1 dB connecting Qingdao and Jinan in Shandong province, China. As far as we know, this is the longest point to point field test of QKD. The secure key rate ($3.45\times 10^{-8}$ per pulse) at 511 km is higher than the absolute PLOB bound, after collecting 4.7 hours data for finite size and fluctuation analysis.

\section*{Protocol}
We adopt the sending-or-not-sending (SNS) protocol~\cite{wang2018sns} with actively odd parity pairing (AOPP)~\cite{xu2019general,jiang2019unconditional} method for post data processing in field experiment of TF-QKD. In this protocol, Alice and Bob use three weak coherent state (WCS) sources in the $X$ basis, $\mu_1$, $\mu_2$, 0,  which are used for the decoy-state analysis. And in the $Z$ basis, Alice and Bob randomly decide sending the WCS pulses $\mu_z$ or not sending, where those pulses are used to extract the final keys. To improve the key rate, we take the odd parity error rejection method through AOPP in the post data processing, which can efficiently deduce the bit-error rate of the raw keys and thus largely improves the key rate. We use the zig-zag approach proposed in Ref.~\cite{jiang2020zigzag} to analyses the finite key effects of AOPP, which assure the high performance of SNS-TF-QKD protocol with AOPP in the finite key size.

In this protocol, Alice and Bob randomly choose the $X$ basis or $Z$ basis with probabilities $1-p_z$ and $p_z$. For the $X$ basis, with probabilities $1-p_1-p_2$, $p_1$ and $p_2$, Alice and Bob randomly prepare and send out pulses with intensities $0$, $\mu_1$ or $\mu_2$. For the $Z$ basis, Alice and Bob randomly decides sending the WCS pulses with intensity $\mu_z$ or not sending with probabilities $\epsilon$ and $1-\epsilon$. For the received pulse pair, if Charlie announces there is only one detector clicks, we call it as a one-detector heralded events. The corresponding bits of one-detector heralded events in $Z$ basis are used to extract the final keys, and the one-detector heralded events in $X$ basis and mismatching basis are used to estimate the yield of single-photon pairs before AOPP. Besides, if the phase of a pulse pair in $X$ basis satisfies the phase post-selection criteria, its corresponding one-detector heralded events can be used to estimate the phase-flip error rate before AOPP. The detail calculation of AOPP is presented in the supplement.

\section*{Experiment}
\begin{figure*}[htb]
	\centering
	\resizebox{15cm}{!}{\includegraphics{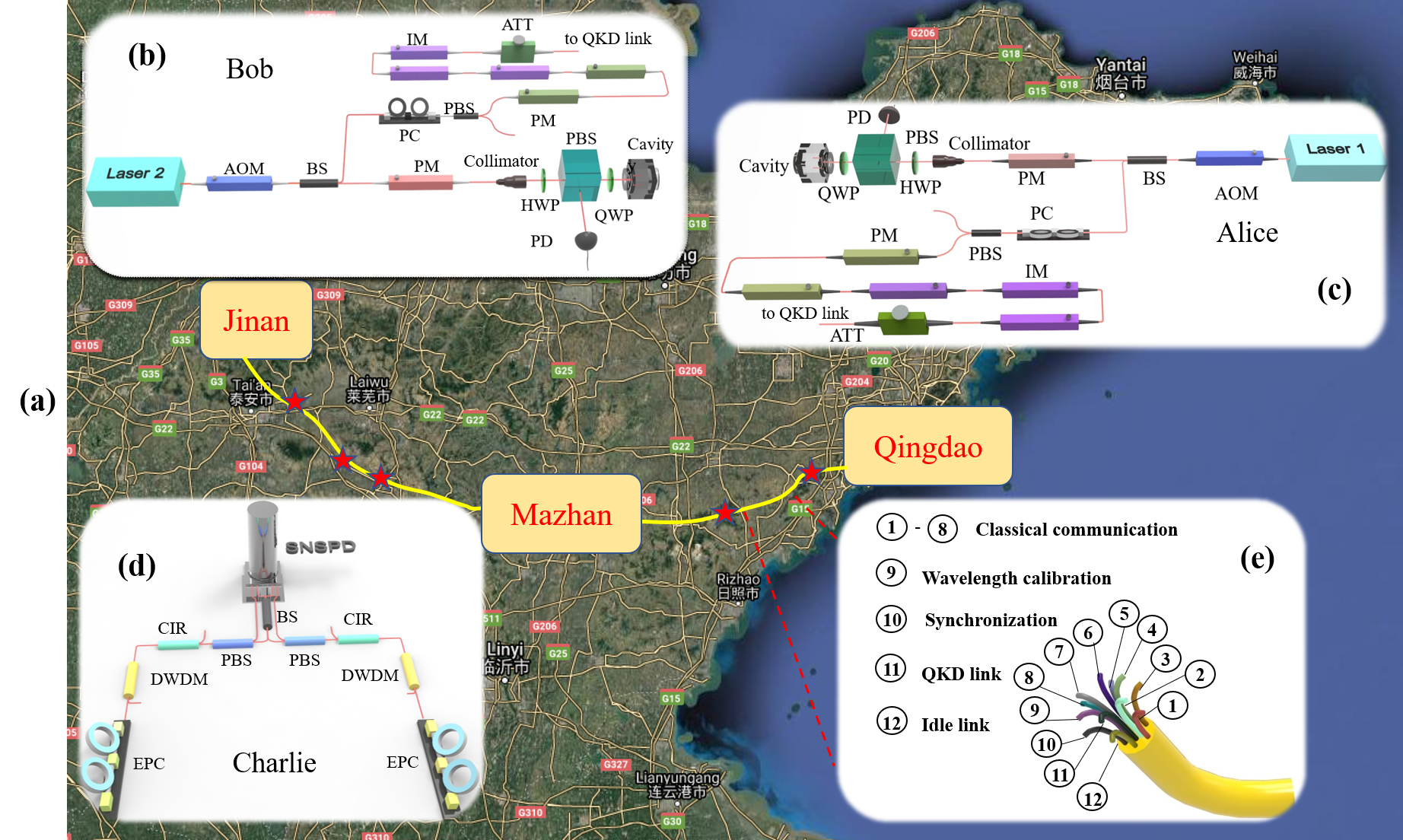}}
	\caption{(\textbf{a}) \baselineskip12pt Satellite image indicating the SNS-TF-QKD experiment with field-deployed fiber between Qingdao (Alice, $N36^{\circ}6'13''$, $E120^{\circ}24'32''$) and Jinan (Bob, $N36^{\circ}36'50''$, $E117^{\circ}6'22''$). Independent lasers are used as light source for  encoding; SNSPDs are used for detection in Mazhan (Charlie, $N36^{\circ}0'19''$, $E118^{\circ}42'35''$) in the middle of the link. (\textbf{b})/(\textbf{c}) Experimental setup of Alice/Bob. In Alice's/Bob's station, a commercial kilo-Hz continuous-wave fiber laser is locked to an ultra-low-expansion (ULE) glass cavity as the light source. AOM: acoustic optical modulator, BS: beam splitter, PM: phase modulator, HWP: half wave plate, PBS: polarization beam splitter, QWP: quarter wave plate, PD: photoelectric detector. Then the frequency locked stable laser is applied to accomplish encoding by two phase modulators (PMs) and three intensity modulators (IMs). Finally the encoded light pulses are attenuated into single photon level with passive attenuator (ATT) and sent to the measurement station. PC: polarization controller. (\textbf{d}) Experimental setup of Charlie. In Charlie's station, two electric polarization controllers (EPCs) are used to real-time feedback the polarization of the pulses from Alice and Bob,  Dense Wavelength Division Multiplexers (DWDMs) to filter the leakage of classical communication in other fibers and the nonlinear scattering from strong reference pulses, and circulators (CIRs) to block the reflection of SNSPDs. (\textbf{e}) Sectional view of a bundle of optical fibers in the field-deployed optical cable. Three of a bundle of 12 fibers are used in this experiment: one for single-photon level signal transmission (QKD link), one for clock synchronization and ``start" signal (Synchronization), and one for wavelength calibration to lock the optical frequency between Alice's and Bob's independent lasers (Wavelength calibration). Besides our experiment, classical communications are running in the other 8 (Classical communication) and one is idle (Idle link) of these 12 fibers.}
	\label{Fig:mapx}
\end{figure*}

We realize TF-QKD in field deployed fiber between two non-adjacent metropolises, Qingdao (Alice, $N36^{\circ}6'13''$, $E120^{\circ}24'32''$) and Jinan (Bob, $N36^{\circ}36'50''$, $E117^{\circ}6'22''$) in Shandong province of China, which are connected by a 430 km long haul fiber network, as shown in Fig~\ref{Fig:mapx}. The measurement station is placed in Mazhan (Charlie, $N36^{\circ}0'19''$, $E118^{\circ}42'35''$) in the middle of the fiber link. The long haul network contains 12 fibers in the cable and mainly used for classical fiber communication test. All the 12 fibers are G654.E ultra-low loss (ULL) fiber with a nominal loss of 0.158 dB/km.

Among the 12 fibers, we have rented 3 for our experiment and the other 9 are still for classical fiber communication of China Unicom Ltd. Our three fibers are utilized to transfer quantum signal, optical synchronizing signal and laser frequency locking signal, respectively. The time frequency dissemination signal is used to lock the two independent lasers' frequency in Jinan and Qingdao. And for the synchronization channel and wavelength locking channel, erbium-doped fiber amplifiers (EDFAs) are used about per 70 km to amplify the signal. While the quantum single doesn't have any relay node.

Similar to our previous TF-QKD experiment in the lab~\cite{chen2020sending,liu2019experimental}, Alice and Bob use independent lasers: the commercial kilo-Hz continuous-wave fiber laser are locked to ultra-low-expansion (ULE) glass cavity in both sides as the light source. The linewidth of both Alice's and Bob's laser is less than 1 Hz. The central wavelength of Alice's laser is set to 1550.12460 nm. We then split the light into two, one part is to transfer quantum signal and the other one is used for wavelength locking. Instead of locking two lasers in a same room, the field experiment needs to lock the two lasers 400 km away. Alice adopts the time frequency dissemination technology~\cite{Predehl441}, and sends the frequency locking laser through the the 84.1 dB (430 km) fiber channel with 6 EDFAs in between. The light interferes with Bob's laser at a photodiode. Based on the beating result, Bob set his laser wavelength with an acoustic-optic modulator (AOM). The relative frequency drift of the two sources is measured to be approximately 0.1~Hz/s, giving an accumulated phase difference of about $\pi/60$ per hour. Therefore, instead of continuing calibrating the wavelength, we only need to calibrate the wavelength difference per hour or two, which is good enough for the experiment. This can eliminate the crosstalk noise from the wavelength locking channel.

\begin{figure*}[htb]
	\centering
	\resizebox{8cm}{!}{\includegraphics{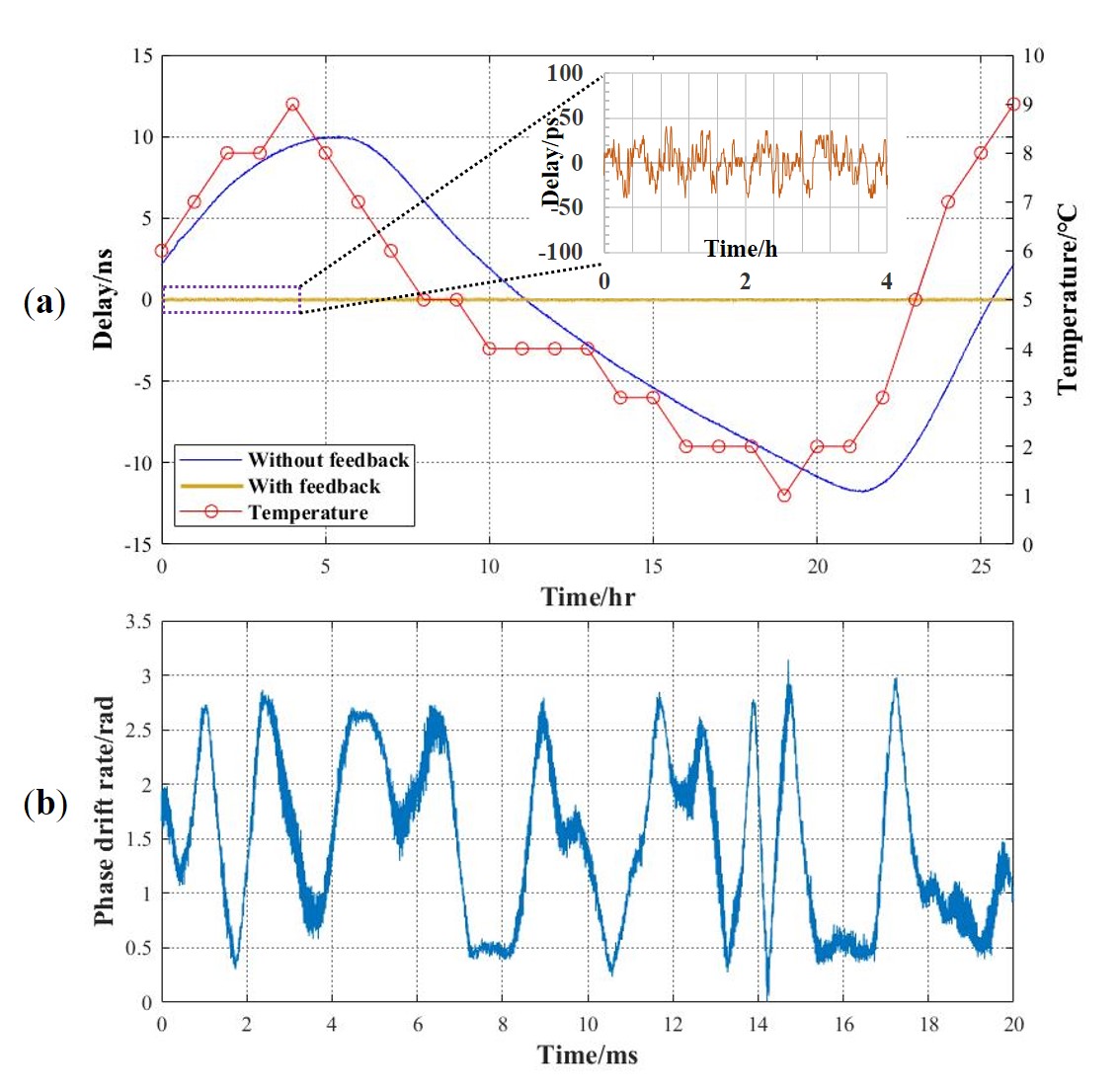}}
	\caption{(\textbf{a}) The signal arrival time  with (orange curve) and without (blue curve) feedback, compared with the temperature change during the measurement (November 24, 2020,  red dot). Insert: Magnified view of the signal arrival time with feedback. (\textbf{b})The phase drift rate is measured as 0.03 rad/us through the field-deployed fiber.}
	\label{Fig:feedback}
\end{figure*}

After locking the wavelength, the fiber fluctuation becomes the main source of phase noise. The ambient temperature change during the day-night cycle affects the effective length of the deployed fiber, and thus the phase and the arrival time of the signal. We monitor the arrival time of the signal by statistic the arriving time of the reference pulse at the PBS's idler port from Alice and Bob. As shown in Fig~\ref{Fig:feedback}(a), without any feedback the arrival time is strongly related to the ambient temperature, which changes up to 20~ns in one day. The arrival time drift may also deteriorate the interference of the signal, by affecting the overlapping between the two pulses.

We utilize two stage procedure to fix it. First, we synchronize Alice and Bob's laser pulses with Charlie's clock~\cite{PhysRevLett.111.130501,caolong2020}. Charlie sends two 250 MHz synchronized laser pulse trains to Alice and Bob through the synchronization fiber channel, respectively. The two laser trains are used to lock the local clock of the signal generators, which then generate signal pulses in both sides with a pulse duration of 280 ps. The measured arrival times in Charlie side are used for feedback signal to adjust the relative delay between Alice's and Bob's synchronization laser trains. With such an arrangement, the fluctuation of the arrival time decrease to ~10 ps, negligible comparing to the signal pulse duration of 280 ps.

After fixing the time drifting , the phase fluctuation is the target for the second stage, which is also larger than the fluctuations in lab, due to environment vibration. We measured a phase rate of 0.03~$rad/\mu s$ through field-deployed fiber, or equivalent to 17 degree in 10 $\mu$s, as shown in Fig~\ref{Fig:feedback}(b). The phase reference pulse and the corresponding phase estimation method~\cite{chen2020sending,liu2019experimental} are used to compensate for this phase drift in post-processing. We time-multiplex the quantum signal together with strong phase reference pulses using intensity modulator. The superconducting nanowire single photon detectors (SNSPD) used in our experiment has a 60~dB dynamic range, which can measure both the quantum signal and strong phase reference signal. According to the measurement result of phase reference signals, we can compensate the phase difference in postprocessing, and then the optical phase sensitive interference is achieved. (See supplementary for details of the experimental setup)

Besides the time drifting and phase fluctuation, the polarization drift in the deployed fiber also ruins the interference. To solve this problem, electric polarization controllers (EPCs) and polarization beam splitters (PBSs) are inserted before the beam splitter in Charlie. The signals from the idler port of the PBS are directed to a SNSPD to monitor the polarization, for Alice and Bob respectively.  The target counting rate is set to 175 kHz for feedback. Without the feedback, the intensity changes 30\% in 4 hours; which is almost stable when the feedback is on. (See Supplementary Material for details).

After fixing all these fluctuations, we also need to filter out the crosstalk noise from adjacent fiber bundles, which is also new comparing to lab experiment. The crosstalk noise count rate from synchronization pulses and classical communication in the fiber bundle is around 486 count per second (cps) and 616 cps for Alice and Bob respectively. We set our QKD wavelength out of the noise spectrum and add a 100~GHz (0.8~nm) Dense Wavelength Division Multiplexing (DWDM) to filter out the noise. After the filter, these crosstalk noises drops to less than 10 cps for both sides.

Besides crosstalk noise counts, there are also noise counts from re-Railey scattering noise~\cite{chen2020sending}, which is caused by the backscattering of the Railey backscattered phase reference light and detector dark counts. We estimate this re-Railey scattering noise to about 8~cps on each single photon detector, based on the detected reference counts. The dark counts for SNSPDs are 4~cps and 6~cps, with detection efficiency of 77\% and 91\% respectively. In total, the noise counts for each detector is about 24~cps. We then further use a 350~ps time filtering to reduce the noise probability per gate, achieving a noise count rate of  $7\times 10^{-9}$.

With all the above improvements, we realized the field test of SNS-TF-QKD with in total 511~km ultra low loss fiber deployed between Jinan and Qingdao. The experimental parameters and results are shown in Tab.~\ref{Tab:Parameters}, and can be found in detail in Supplementary Materials. With an effective system frequency of 100 MHz, we sent $1.679\times 10^{12}$ pulses in approximately 4.7 hours. A total of $4.987\times10^6$ valid detections are recorded, where one and only one detector clicks. The raw bit-flip error rate before AOPP is 24.39\%, which drops to 0.431\% after AOPP. The survived bits after AOPP decreases to 576130. These values are in agreement with the theoretically expected values very well, which are 0.426\% and 576295 respectively.

The secure key rate considering data fluctuation and the finite data size effect is calculated following the theory~\cite{jiang2019unconditional}:

\begin{equation}\label{eq:KeyRate}
\begin{split}
 R=&\frac{1}{N_t}\{ n_1^\prime[1-H(e_1^{ph^\prime})]-fn_t^\prime H(E_Z)\}\\
 &-2\log_2\frac{2}{\varepsilon_{cor}}-2\log_2\frac{1}{\sqrt{2}\varepsilon_{PA}\hat{\varepsilon}}\},\\
\end{split}
\end{equation}

where $R$ is the final key rate, $n_1^\prime$, $e_1^{ph}$,  $n_t^\prime$ and $E_Z$ are the number of untagged-bits, the phase-flip error rate, the number of survived bits, and the bit-flip error rate of untagged-bits after AOPP, $f$ the error correction efficiency where we set to $f=1.1$, $N_{t}$ the total number of signal pulses, $\varepsilon_{cor}=1\times 10^{-10}$ the failure probability of error correction process, $\varepsilon_{PA}=1\times 10^{-10}$ the failure probability of privacy amplification process, $\hat{\varepsilon}=1\times 10^{-10}$ the coefficient of the chain rules of smooth min- and max- entropy.
\linespread{1.5}
\begin{table*}[htb]
\centering
  \caption{Experimental parameters and results. The parameters $\mu_1$, $\mu_2$ and $\mu_z$ are the intensities of the sent pulses of weak decoy state, strong decoy state and signal state, $p_1$ and $p_2$ are the probabilities of weak decoy state and strong decoy in $X$ basis, $p_z$ is the probability of $Z$ basis, $\epsilon$ is the sending probability in $Z$ basis.}
\begin{tabular}{c||cc}
\hline
\hline
$\mu_1$  & 0.100 \\
$\mu_2$  & 0.298\\
$\mu_z$  & 0.422 \\
$p_1$ & 0.846 \\
$p_2$ & 0.076\\
$p_z$ & 0.735 \\
$\epsilon$ & 0.269 \\
\hline
$n_t^\prime$  & 576130 \\
$n_1^\prime$  & 219136  \\
$e_1^{ph}$   & 16.06\%  \\
$E_Z$   & 0.43\%\\
\hline
$N_{t}$	 & $1.679\times 10^{12}$ \\
Valid detections &$4.987\times10^{6}$\\
Key rate			& $3.45\times10^{-8}$ \\
\hline
\end{tabular}
\label{Tab:Parameters}
\end{table*}

\begin{figure*}[htb]
	\centering
	\resizebox{8cm}{!}
	{\includegraphics{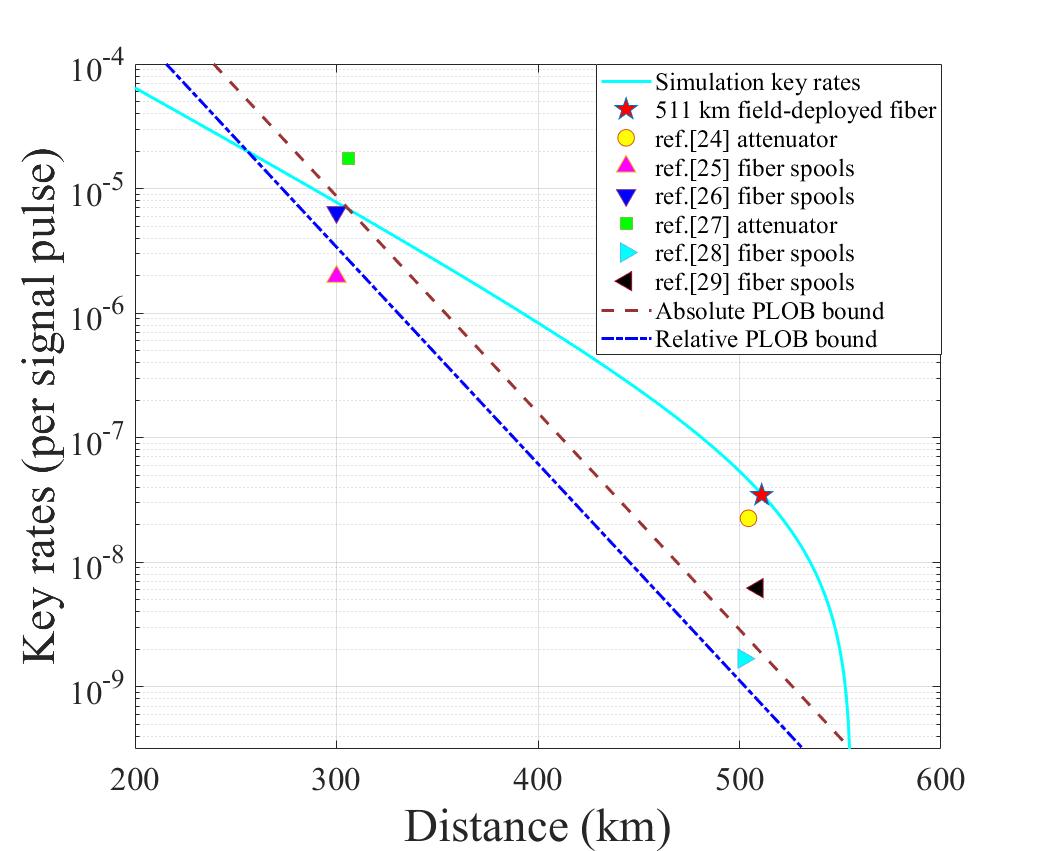}}
	\caption{\textbf{Secure key rates of the SNS-TF-QKD experiment.} The red star indicates the experimental result over  511 km field-deployed fibers, with the secure key rate of $R=3.45\times10^{-8}$. The yellow circle and the green square experimental result of~ref.[24] and~ref.[27], respectively, with attenuators simulating the channel loss. The four triangles indicate the experimental results of~ref.[25],~ref.[26],~ref.[28] and~ref.[29], with in lab fiber spools. The cyan curve is the simulation result with the experimental parameters. The blue dotted line and brown dashed line shows the relative and absolute PLOB bound~\cite{PLOB2017}.}
	\label{Fig:511km}
\end{figure*}

The secure key rate is calculated as $R=3.45\times10^{-8}$, which is about 3.45 bps considering our system frequency. We plot this value and the theoretical simulation in Fig.~\ref{Fig:511km}, where we also plot recent TF-QKD experimental results as a comparison. With the improvement in both theory and experiment, our secure key rate is more than 10 times higher than the absolute PLOB bound. The secure key rate is also 3 orders higher than previous field test MDI-QKD experiment~\cite{tang2016measurement}, if considering linear decreasing and assuming no detector dark count. This result demonstrates the capability of the SNS-TF-QKD protocol with AOPP in practical environment with field deployed fibers.

\section*{Conclusion}
In conclusion, we experimentally demonstrated SNS-TF-QKD with two independent laser sources, in 511~km long haul field-deployed fibers between the city of Jinan and Qingdao. The generated secure key rate is an order higher than the absolute key rate limit of the repeater-less QKD protocol and 3 orders of magnitudes greater than what is expected if the previous QKD field test system over the same length were applied. Our work proves the feasibility of TF-QKD in a practical environment and paves the way towards long haul fiber quantum  network. The techniques developed in the work will find immediate applications in more general applications like in quantum repeater~\cite{dlcz}, and phase-based architecture for the quantum internet~\cite{wehnerquantum2018}.

\clearpage
\textbf{\subsection*{Acknowledgments}}
This work was supported by the National Key R\&D Program of China (Grants No. 2017YFA0303900, 2017YFA0304000, 2020YFA0309800), the National Natural Science Foundation of China, the Chinese Academy of Sciences (CAS), Shanghai Municipal Science and Technology Major Project (Grant No.2019SHZDZX01), Key R\&D Plan of Shandong Province (Grant No. 2019JZZY010205, 2020CXGC010105), the Taishan Scholar Program of Shandong Province, and Anhui Initiative in Quantum Information Technologies.

\clearpage
\begin{appendix}
\section*{The SNS-TF-QKD protocol with odd-parity error rejection}

The 4-intensity SNS-TF-QKD protocol is used to perform the experiment, and the theory of actively odd-parity paring (AOPP) with finite key effects is used to get high key rate.

In the whole protocol, Alice and Bob send $N$ pulse pairs to Charlie, and Charlie is supposed to announce an one-detector heralded event if his received pulse pair cause only one detector clicks.
At each time, Alice (Bob) randomly chooses the $X$ basis with probability $1-p_{z}$, or a $Z$ basis with probability $p_{z}$. If the $Z$ basis is chosen, with probability $\epsilon$, Alice (Bob) randomly prepares a phase-randomized weak coherent state (WCS) pulse with intensity $\mu_{z}$, and denote it as bit $1$ ($0$); with probability $1-\epsilon$, Alice (Bob) prepares a vacuum pulse, and denote it as bit $0$ ($1$). If the $X$ basis is chosen, Alice (Bob) randomly prepares a vacuum pulse or coherent state pulse $\ket{e^{i\theta_{A1}}\sqrt{\mu_{1}}}$ or $\ket{e^{i\theta_{A2}}\sqrt{\mu_{2}}}$ ($\ket{e^{i\theta_{B1}}\sqrt{\mu_{1}}}$ or $\ket{e^{i\theta_{B2}}\sqrt{\mu_{2}}}$) with probabilities $p_0=1-p_{1}-p_2$, $p_1$ and $p_{2}$, respectively, where $\theta_{A1},\theta_{A2},\theta_{B1}$ and $\theta_{B2}$ are different in different windows, and are random in $[0,2\pi)$.

A time window that both Alice and Bob choose the $Z$ basis is called a $Z$ window. One-detector-heralded events in $Z$ windows are called effective events. Alice and Bob get two $n_t$-bit strings formed by the corresponding bits of effective events of the $Z$ windows. For a time window that both the intensities of Alice's  and Bob's WCS pulse is $\mu_{1}$, and if the phases of the WCS pulses satisfy
\begin{equation}
1-\vert \cos(\theta_{A1}-\theta_{B1}-\psi_{AB})\vert\le \lambda,
\end{equation}
it is called an $X$ window. Here, $\psi_{AB}$ can take an arbitrary value, which can be different from time to time as Alice and Bob like, so as to obtain a satisfactory key rate for the protocol~\cite{liu2019experimental}. $\lambda$ is a positive value close to $0$, and would be optimized to obtain the highest key rate. One-detector-heralded events in $X$ windows are called effective events.

The data of one-detector heralded events are used to decoy method analysis. To clearly show the calculation method, we denote the vacuum source, the WCS source with intensity $\mu_{1},\mu_{2}$, and $\mu_z$  of Alice (Bob) by $o,x,y$ and $z$ ($o^\prime,x^\prime,y^\prime$, and $z^\prime$). We denote the number of pulse pairs of source $\kappa\zeta(\kappa=o,x,y,z;\zeta=o^\prime,x^\prime,y^\prime,z^{\prime})$ sent out in the whole protocol by $N_{\kappa\zeta}$, and the total number of one-detector heralded events of source $\kappa\zeta$ by $n_{\kappa\zeta}$. We define the counting rate of source $\kappa\zeta$ by $S_{\kappa\zeta}=n_{\kappa\zeta}/N_{\kappa\zeta}$, and the corresponding expected value by $\mean{S_{\kappa\zeta}}$.

Then we can use the decoy-state method to calculate the lower bounds of the expected values of the counting rate of single-photon states $\oprod{01}{01}$ and $\oprod{10}{10}$, which are~\cite{hu2019general}
\begin{align}
\label{s01mean}\mean{\underline{s_{01}}}&= \frac{\mu_{2}^{2}e^{\mu_{1}}\mean{S_{ox^\prime}}-\mu_{1}^{2}e^{\mu_{2}}\mean{S_{oy^\prime}}-(\mu_{2}^{2}-\mu_{1}^{2})\mean{S_{oo^\prime}}}{\mu_{2}\mu_{1}(\mu_{2}-\mu_{1})},\\
\mean{\underline{s_{10}}}&= \frac{\mu_{2}^{2}e^{\mu_{1}}\mean{S_{xo^\prime}}-\mu_{1}^{2}e^{\mu_{2}}\mean{S_{yo^\prime}}-(\mu_{2}^{2}-\mu_{1}^{2})\mean{S_{oo^\prime}}}{\mu_{2}\mu_{1}(\mu_{2}-\mu_{1})}.
\end{align}
Then we can get the lower bound of the expected value of the counting rate of untagged photons
\begin{equation}
\mean{\underline{s_1}}=\frac{1}{2}(\mean{\underline{s_{10}}}+\mean{\underline{s_{01}}}),
\end{equation}
and the lower bound of the expected value of the untagged bits $1$, $\mean{\underline{n_{10}}}$, and untagged bits $0$, $\mean{\underline{n_{01}}}$
\begin{align}
\mean{\underline{n_{10}}}=Np_{z}^2\epsilon(1-\epsilon)\mu_{z}e^{-\mu_{z}}\mean{\underline{s_{10}}},\\
\mean{\underline{n_{01}}}=Np_{z}^2\epsilon(1-\epsilon)\mu_{z}e^{-\mu_{z}}\mean{\underline{s_{01}}}.
\end{align}

We denote the number of total pulses sent out in the $X$ windows by $N_{X}$, and the number of corresponding error events by $m_{X}$, then we have the error counting rate of $X$ windows
\begin{equation}
T_{X}=\frac{m_{X}}{N_{X}}.
\end{equation}
Then we have
\begin{equation}\label{e1}
\mean{\overline{e_1^{ph}}}=\frac{\mean{T_{X}}-e^{-{2\mu_{1}}\mean{S_{oo^\prime}}/2}}{2e^{-2\mu_1}\mu_1\mean{\underline{s_1}}},
\end{equation}
where $\mean{T_{X}}$ is the expected value of $T_{X}$.

With values above, we can calculate the lower bound of untagged bits and phase-flip error rate after AOPP, $n_1^\prime$ and $e_{1}^{\prime ph}$ by the method proposed in Ref.~\cite{jiang2020zigzag}. We have the related formulas as follows.
\begin{subequations}
\begin{align}
&u=\frac{n_g}{2n_{odd}},\quad \underline{n_{1}}=\varphi^L(\mean{\underline{n_{10}}}+\mean{\underline{n_{01}}}),\\
&\underline{n_{10}}=\varphi^L(\mean{\underline{n_{10}}}),\underline{n_{01}}=\varphi^L(\mean{\underline{n_{01}}}),\\
&n=\varphi^L\left(\frac{{\underline{n_{1}}}}{n_t}\frac{{\underline{n_{1}}}}{n_t}\frac{un_t}{2}\right),\\
&k=u\underline{n_{1}}-2n,\\
&r=\frac{2n+k}{k}\ln\frac{3k^2}{\varepsilon(r,k)},\\
&\bar{M}=\varphi^U(2n\mean{\overline{e_1^{ph}}}),\\
&\mean{e_{\tau}}=\frac{\bar{M}}{2n-r}\\
&\bar{M}_s=\varphi^U[(n-r)\mean{e_{\tau}}(1-\mean{e_{\tau}})]+r,
\end{align}
\end{subequations}
where $n_g$ is the number of pair if Alice and Bob perform AOPP to their raw keys, ${n_{odd}}$ is the number of pairs with odd-parity if Bob randomly groups all the bits in $Z_B$ two by two, and $n_g$ and $n_{odd}$ are observed values, $\varepsilon(r,k)=10e^{-10}$ is the trace distance while using the exponential de Finetti's representation theorem, and $\varphi^U(x),\varphi^L(x)$ are the upper and lower bounds while using Chernoff bound~\cite{chernoff1952measure} to estimate the real values according to the expected values.

Finally, we can get the key rate $R$ by
\begin{equation}\label{r2}
\begin{split}
R=&\frac{1}{N}\left\{n_1^\prime[1-h(e_{1}^{\prime ph})]-fn_t^\prime h(E^\prime)-2\log_2{\frac{2}{\varepsilon_{cor}}}\right.\\
&\left.-2\log_2{\frac{1}{\sqrt{2}\varepsilon_{PA}\hat{\varepsilon}}}\right\}.
\end{split}
\end{equation}
where $n_1^\prime$ is the number of the untagged bits after AOPP, $e_{1}^{\prime ph}$ is the bit flip error rate of untagged bits after AOPP, $h(x)=-x\log_2x-(1-x)\log_2(1-x)$ is the Shannon entropy, $E^\prime$ is the bit-flip error rate of the remaining bits after AOPP, $\varepsilon_{cor}$ is the failure probability of error correction, $\varepsilon_{PA}$ is the failure probability of privacy amplification, and $\hat{\varepsilon}$ is the coefficient while using the chain rules of smooth min- and max- entropy~\cite{vitanov2013chain}.

\section*{Setup of the experiment}

\begin{figure*}[tbh]
	\centering
	\resizebox{14cm}{!}{\includegraphics{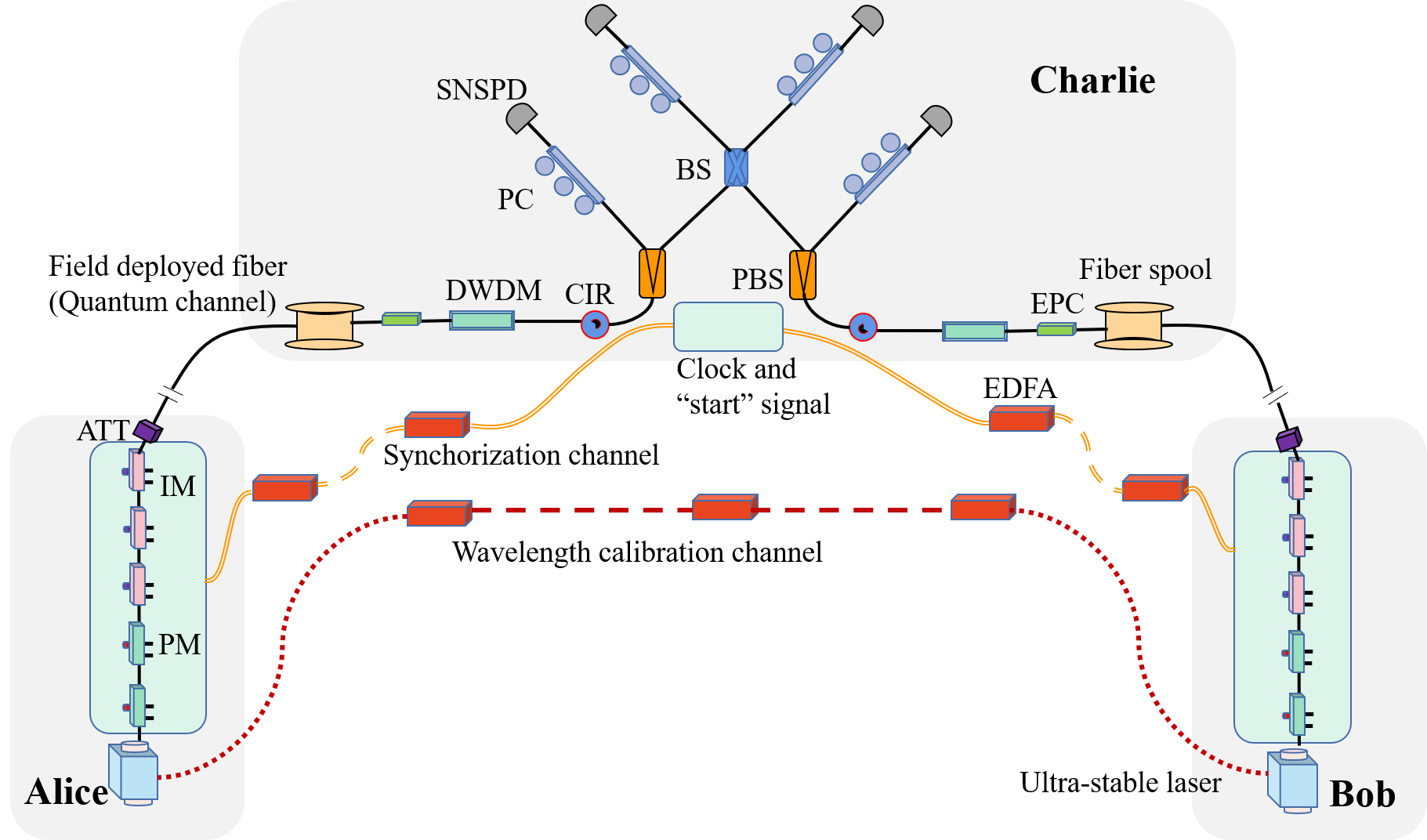}}
	\caption{\textbf{chematic of our experimental setup.} \baselineskip12pt Alice and Bob respectively adopt a ultra stable continuous wave (CW) laser with a linewidth less than 1~Hz by using the technique of Pound-DreverHall (PDH)~\cite{drever1983laser,pound1946electronic}. These light sources are then modulated by two phase modulators (PMs) and three intensity modulators (IMs) for phase randomization, reference and decoy intensity modulation. The pulses are then attenuated by an attenuator (ATT) and sent out via quantum fiber channel to Charlie. At Charlie's station, the encoded pulses from Alice and Bob are adjusted with electric polarization controllers (EPCs) and polarized with polarization beam splitters (PBSs), filtered with Dense Wavelength Division Multiplexings (DWDMs) and circulators (CIRs), and finally interfered at a beam splitter (BS). The pulses are detected by superconducting nanowire single-photon detectors (SNSPDs), PC: polarization controller. Besides the Quantum fiber channel, two additional fibers are used in this experiment: one for clock synchronization and ``start" signal, and one for wavelength calibration to lock the optical frequency between Alice's and Bob's independent lasers. For about every 70~km, EDFAs are used to maintain the optical power of these classical signals.}
	\label{Fig:setup}
\end{figure*}

Since the field experimental conditions are different from that in the laboratory where the channel can be well controlled, the real-life conditions have to be considered now. For example, the wavelength locking of physical separated lasers, the additional crosstalk noise from adjacent fibers, the larger phase fluctuation and even the drifting in the signal arrival time. Considering these challenges, we have fully upgraded our whole system in experimental side. To implement the field SNS-TF-QKD above, we make our experimental setup as shown in Fig.~\ref{Fig:setup}. Among the 12 fibers in one cable, we have rented 3 for our experiment: one for single-photon level signal transmission, one for clock synchronization and ``start" signal distribution, and one for wavelength calibration to lock the optical frequency between Alice's and Bob's independent lasers. For about every 70~km, EDFAs are used to maintain the optical power of these classical signals. Besides our experiment, the other 9 are still for classical fiber communication of China Unicom Ltd.

\clearpage

\subsection*{Stable independent light sources}

Similar to our previous TF-QKD experiment in the lab~\cite{chen2020sending,liu2019experimental}, Alice and Bob respectively adopt a ultra stable continuous wave (CW) laser with a linewidth less than 1~Hz as the light source. First, they use a commercial continuous wave laser (NKT E15) with a linewidth of about 1~kHz at the center wavelength of 1550.12~nm as the seed laser. Then Alice locks her seed laser to a 5-cm-long ultra-stable cavity with a finesse around 400000, and Bob locks his seed laser to a 10-cm-long ultra-stable cavity with a finesse around 400000, using the Pound-DreverHall (PDH) technique~\cite{drever1983laser,pound1946electronic}, to suppress their line-width from about 1~kHz to less than 1~Hz. Since the ultra-stable cavity is machined by ultra-low-expansion (ULE) glass, the resonant frequency of each cavity is different, after the seed laser is respectively locked to their cavity, finally, they additionally need to calibrate the frequency/wavelength difference of the two stable laser sources. Compared to the experiments in the laboratory~\cite{chen2020sending,liu2019experimental}, we change the way the twin-field light sources works to decrease the crosstalk noise. We calibrate the frequency difference of the two independent light sources intermittently instead of continuous frequency locking. We first calibrate the wavelength difference of the two independent light sources before the experiment, with strong CW reference light in the fiber; Then when the experiment is running, the strong reference is blocked, and the wavelength between the two light sources are left to free drifting. With stable light sources, the wavelength difference will be small for the experiment between the wavelength calibration cycles.

\begin{figure*}[htb]
	\centering
	\resizebox{9cm}{!}
	{\includegraphics{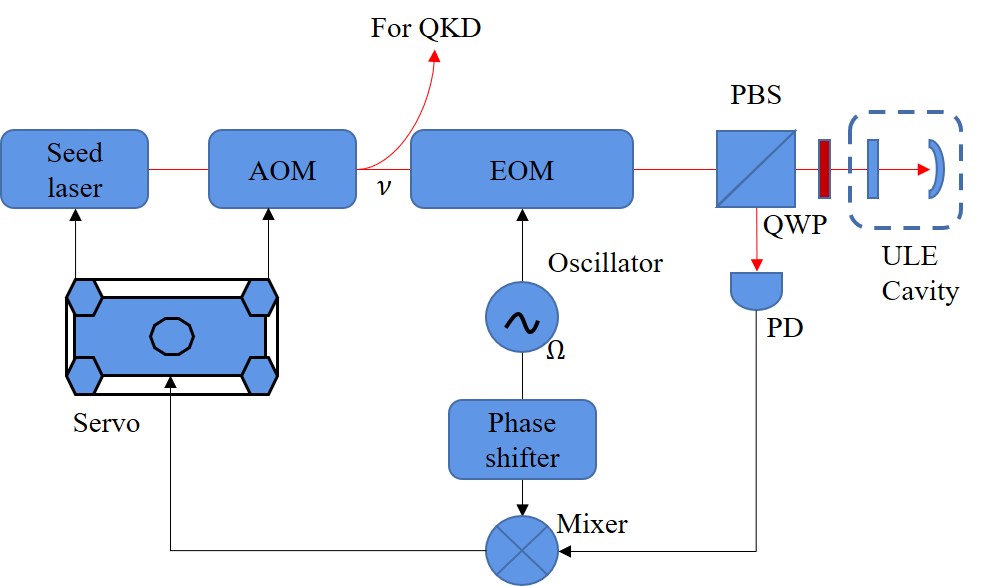}}
	\caption{\textbf{Scheme of the PDH technique.}}
\label{Fig:PDH}
\end{figure*}

Here we give a brief introduction on the PDH method. As shown in Fig.~\ref{Fig:PDH}, after frequency modulated by an acoustic optical modulator (AOM), the seed laser with a frequency of $\nu$ is phase modulated by an electro-optical modulator (EOM) with a radio-frequency (RF) of $\frac{\Omega}{2\pi}$ coupling in the ULE ultra-stable cavity through a polarization beam splitter (PBS) and a quarter-wave plate (QWP), then the input of the cavity contains optical signals at frequencies of $\nu$ - $\frac{\Omega}{2\pi}$ , $\nu$ and $\nu$ + $\frac{\Omega}{2\pi}$. The signals at frequencies of $\nu$ - $\frac{\Omega}{2\pi}$ and $\nu$ + $\frac{\Omega}{2\pi}$ are sidebands of the modulation. Then detected by a photoelectric detector (PD), the reflected carrier wave is the beat-note signal between the carrier and sidebands. If the carrier frequency $\nu$ of the seed laser is close to a resonant frequency of the cavity, the reflected carrier wave can be regarded as a combination of the backward leakage of cavity-stored signal and the reflected incident signal. Further with the reference signal from the oscillator, the corresponding error signal is obtained by demodulating the beat-note by a mixer. The phase shifter is required to set the phase relationship between the demodulation and detected signals to be in phase, leading to a maximum demodulation efficiency. Thus based on the error signal, we use two proportional integral servo controllers to realize the feedback. The slow feedback with a bandwidth of about 6~kHz is accomplished by adjusting the piezoelectric (PZT) of the seed laser, and the fast feedback with a bandwidth of about 100~kHz by adjusting the carrier frequency of the AOM.

\subsection*{The encoding details}

Based on these two high-performance narrow line-width coherent light sources, Alice and Bob individually implement the encoding by applying two phase modulators (PMs) to encode the light to 16 different phase slices and three intensity modulators (IMs) to 5 different intensities respectively. With each basic period of 1 $\mu$s time sequence, the first 500~ns they send 100 signal pulses with 4 different random intensities and 16 phase values are respectively used as the signal state pluses, strong decoy state pluses, weak decoy state pluses, and vacuum state pluses, each with a 280~ps pulse duration and 4.72~ns interval, the following 400~ns they send 4 strong phase reference pulses with a same intensity and determined phase values, each with a 100~ns pulse duration, the last 100~ns they send nothing as the recovery time for the superconducting nanowire single-photon detectors (SNSPDs).

For the phase encoding, during the basic period of 1 $\mu$s, Alice (Bob) randomly modulates the phase to $\theta_A$ ($\theta_B$) on 100 signal pulses with a 5~ns duration in the first 500~ns, $\theta_A$, $\theta_B\in\{0, \pi/8, 2\pi/8 ... 15\pi/8\}$, in the next 400~ns, Alice orderly modulates the phase to 0, $\pi/2$, $\pi$, $3\pi/2$, while Bob modulates the phase to $\pi$ on 4 reference pulses with a each 100~ns duration, and then they both modulate the phase to $\pi$ on vacuum state pulses in the last 100~ns. Following the phase modulation, the two users modulate the light to 5 different intensities with three IMs. During the basic period of 1 $\mu$s, Alice and Bob randomly modulate the pulse intensity to $\mu_x$, $\mu_x\in \{\mu_z, \mu_1, \mu_2, 0 \}$ on 100 signal pulses with a 280~ps duration and 4.72 ns interval in the first 500~ns, $\mu_z$ is the intensity of the signal state, $\mu_1$ is the intensity of the weak decoy state, $\mu_2$ is the intensity of the strong decoy state, and $0$ is the vacuum state, in the next 400~ns, they modulate the intensity to $\mu_{ref}$ on 4 reference pulses with a each 100~ns duration, finally they modulate the intensity to $0$ on vacuum state pulses in the last 100~ns.

With pre-generated quantum numbers, all the modulation signals are generated by a high-speed AWG (Tektronix, AWG70002A) with a maximum output of 500 mV and amplified by the electronic amplifiers, thus, in a basic period, the encoding waveform of the phase and intensity modulators would randomly generate different durations and amplitudes patterns, especially the full wave driving voltage of the PM is approximate to 9 V peak-peak value, normal RF amplifiers are difficult to maintain without distortion after amplifying the waveform with a such high gain, however, the distortion of the driving waveform of the PM would inevitably contribute to QBER in $X$-basis and that of the IM would decrease the stability of the encoding, therefore by applying 5 linear voltage amplifiers (LTC6433) with a maximum output of 5.2 V peak-peak to drive the two PMs and three IMs, Alice and Bob can maintain low modulation error rate and high encoding stability. Additionally, Alice and Bob insert an active bias point feedback to the first IM which encodes the decoy state and place all the PMs and IMs into a thick foam box to reduce the fluctuations of the ambient temperature to improve the stability of the encoding. After that the signals from both sides are attenuated into single photon level with passive attenuators and sent to the measurement station (Charlie) through a total channel length of 511 km with a loss of 89.1 dB.

\subsection*{The field fiber links}

In the experiment, Alice and Bob are placed in two non-adjacent metropolises, Qingdao (Alice, $N36^{\circ}6'13''$, $E120^{\circ}24'32''$) and Jinan (Bob, $N36^{\circ}36'50''$, $E117^{\circ}6'22''$) which are separated by other two metropolises, Zibo and Weifang, while Charlie in Mazhan (Charlie, $N36^{\circ}0'19''$, $E118^{\circ}42'35''$) in the middle of the fiber link, in Shandong province, China. We in total used three fiber links, the QKD link, synchronization link and wavelength calibration link. They are all in a same bundle, meanwhile there are one idle fiber link and 8 running fiber links of classical communication in it. Based on the QKD link, Alice and Bob spatia separately apply a high speed arbitrary waveform generator (AWG) to randomly generate encoding waveforms, then amplify the waveforms by electronics amplifiers to drive the phase modulators (PMs) and intensity modulators (IMs) to modulate their stable CW laser, ultimately send the modulated pulses to Charlie for interference.

A total channel length of 511 km ultra-low-loss fiber (G654.E ULL) of the QKD link connecting Alice, Charlie and Bob, which consists of 430~km buried fiber with a loss of 76~dB and 81~km ultra-low loss optical fiber spool with a loss of 13.1~dB. The physical separation distance is more than 300~km between Alice and Bob. The link from Alice (Bob) to Charlie consists of 207.6~km (222.7~km) buried fiber with a loss of 35.5~dB (40.5~dB) and 55.5~km (25.5~km) ultra-low loss fiber spool with a loss of 9~dB (4.1~dB). The total loss of the ultra-low loss fiber link is 89.1~dB (Alice to Charlie 44.5~dB, Bob to Charlie 44.6~dB), which is around 0.174~dB/km on average. The excess loss than the nominal loss of G654.E fiber (0.158~dB/km) is attributed to the fiber connections at the nodes along the link. There are two (three) nodes along the link between Alice (Bob) and Charlie, the corresponding lengths between nodes are shown in Tab.~\ref{Tab:nodes}.

\begin{table*}[htb]
\centering
  \caption{The fiber length between nodes.}
\begin{tabular}{c|c}
\hline
\hline
Parameters & Fiber Length  \\
\hline
Qingdao (Alice) to Huangshan  & 68~km \\
Huangshan to Zhucheng         & 68.3~km  \\
Zhucheng to Mazhan (Charlie)  & 71.3~km  \\
Fiber spool (Alice's side)    & 55.5~km \\
\hline
Jinan (Bob) to Bogeyan  & 71.4~km  \\
Bogeyan to Yuanshan         & 28.9~km  \\
Yuanshan to Yiyuan  & 62~km \\
YiYuan to Mazhan (Charlie)  & 60.4~km  \\
Fiber spool (Bob's side)    & 25.5~km \\
\hline
\end{tabular}
\label{Tab:nodes}
\end{table*}

Through the synchronization link (430~km length, 82.6~dB loss), Charlie distributes synchronous clocks and the system start signals to Alice and Bob to synchronize and trigger their AWG, and to synchronize his own high speed multichannel time tagger locally. At the station of Charlie, an AWG (Tektronix, AWG5202) is used to drive a multi-channel external modulation laser module to generate 250~MHz pulses train with a wavelength of 1548.11~nm as the synchronous clock, and every time at the beginning of the experiment, to originate 100~kHz pulses train with a wavelength of 1560.20~nm as system start signal. In the following the synchronous clock and system start signal are respectively sent to Alice and Bob, and 5 EDFAs are inserted in the link (EDFAs are placed in about every 70~km, Bogeyan, Yuanshan, Yiyuan, Zhucheng, Huangshan) to relay them. Accordingly, at the station of Alice (Bob), a DWDM and two photoelectric detectors are used to extract the synchronous clock and system start signals.

\begin{figure*}[htb]
	\centering
	\resizebox{9cm}{!}
	{\includegraphics{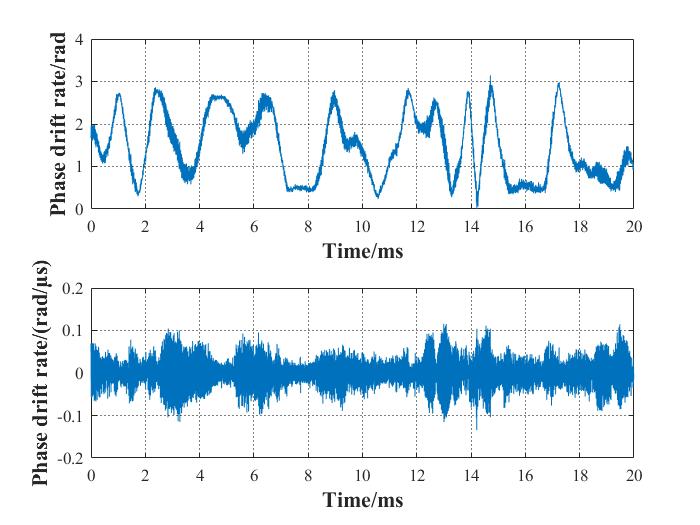}}
	\caption{\textbf{The relative phase drift of field fiber links.}}
\label{Fig:interference}
\end{figure*}

Additionally, by using the wavelength calibration link, Alice send the wavelength reference laser to Bob for calibrating the wavelength difference of the two independent laser sources. The total length of the wavelength calibration link is 430~km, 6 erbium-doped fiber amplifiers (EDFAs) are used in between to compensate a loss of 84.1~dB. At the station of Bob, when he receives the wavelength reference laser from Alice, he divides parts of his light with a intensity of about 500~$\mu$w interferes with the received reference light at a photodiode. Based on the beating result, Bob set his laser wavelength to the same as Alice's with an acoustic-optic modulator (AOM). After the frequency/wavelength of the two lasers is calibrated to be same, we measured the relative phase drift of the twin-field (phase drift angle and the phase drift rate) with field fiber links between Alice and Bob, as is shown in Fig.~\ref{Fig:interference}, the corresponding phase drift rate follows a Gaussian distribution with a standard deviation of $0.03 rad\cdot \mu s^{-1}$. In the experiment, we applied accumulating of 10 basic periods (10 $\mu$s) of the reference detections for the relative phase estimation, the corresponding relative phase drift is around 0.3 rad (or 17 degree) which contributes to an error is less than $4\%$ when Alice and Bob sent the same phase.

\begin{figure*}[htb]
	\centering
	\resizebox{9cm}{!}
	{\includegraphics{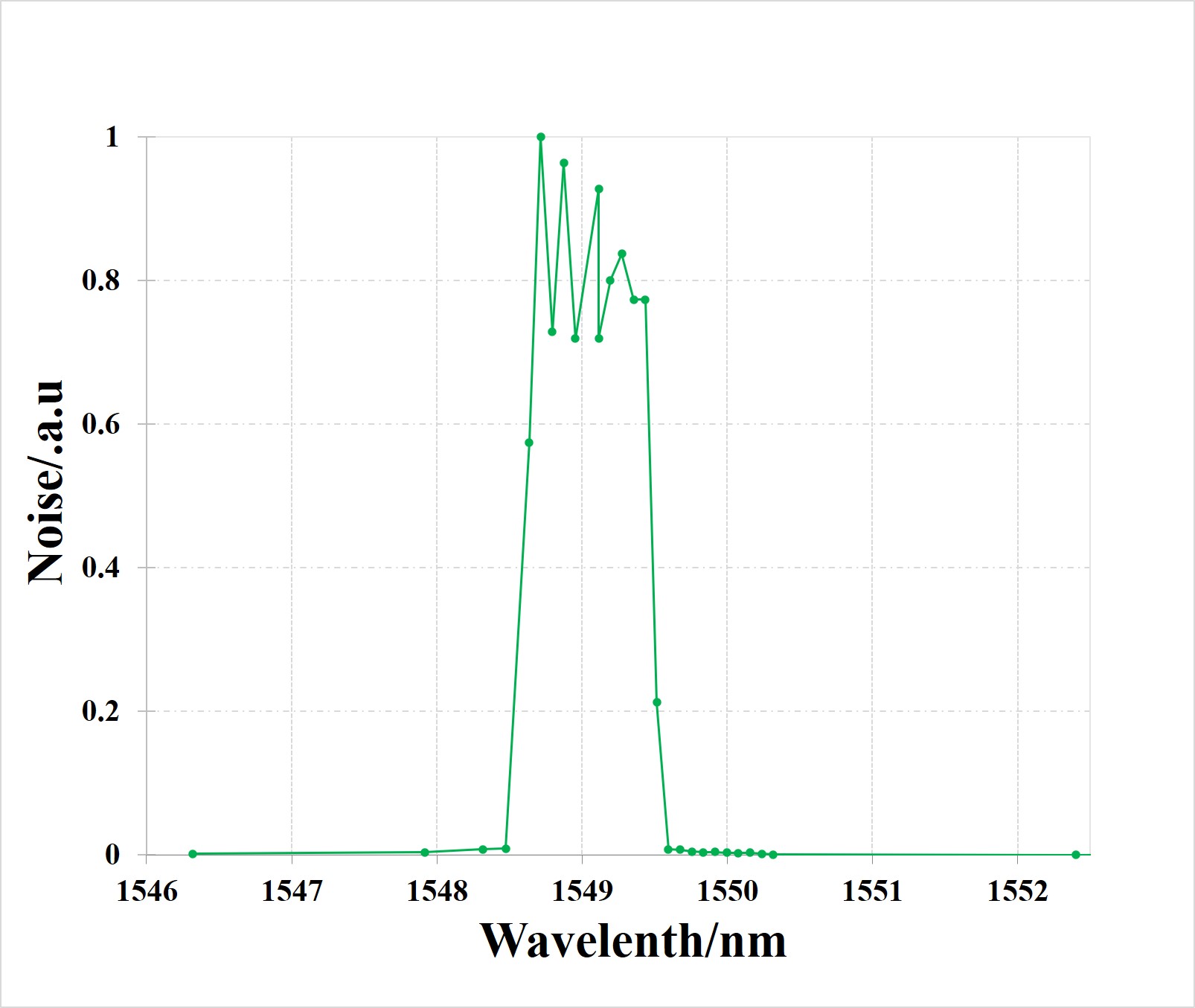}}
	\caption{\textbf{The distribution of the crosstalk from the classical communication.}}
	\label{Fig:noise}
\end{figure*}

Since the synchronization link, wavelength calibration link and other 8 running fiber links of classical communication in the same bundle with the QKD link. The signals running in these links would leak into the QKD link as noise. We measured the spectral distribution of the crosstalk from the 8 running fiber links of classical communication, which is shown in Fig.~\ref{Fig:noise}, the crosstalk is mainly distributed between 1548.5~nm and 1549.5~nm. Therefore we set a wavelength of 1550.12460 nm of the QKD sources to keep away from the main distribution range of crosstalk from the 8 running fiber links of classical communication, and insert a 100~GHz DWDM (1550.12~nm) in measurement station to filter the noise. Meanwhile the leakage of the synchronization pulses is also filtered out with the DWDM. In addition, another important source of noise is the leaked wavelength calibration light. As the wavelength of the calibration light is the same as the signal, we cannot filter them with the DWDM. To reduce this noise, we adopt the time-multiplexing method. We first calibrate the wavelength difference of the two independent light sources before the experiment, with strong CW reference light in the fiber; Then when the experiment is running, the strong reference is blocked, and the wavelength between the two light sources are left to free drifting. With stable light sources, the wavelength difference will be small for the experiment between the wavelength calibration cycles. We usually calibrate it about every one or two hours. Finally, as is shown in Tab.~\ref{Tab:noise}, we tested all the noises above and the dark count of the SNSPDs, and estimate the noises from re-Railey scattering noise~\cite{chen2020sending}. Sum all the above noise sources together, we achieve about 24~cps (count per second) noise counts for each detector. We further use a 350 ps narrow gate, to reduce the noise probability per gate, achieving a dark count rate of $7 \times 10^{-9}$.

\begin{table*}[htb]
\centering
  \caption{The noise in the system.}
\begin{tabular}{c|cc}
\hline
\hline
Noise & Counts  \\
\hline
Classic communication leakage from Alice (direct output)  & 486~cps \\
Classic communication leakage from Bob (direct output)         & 616~cps\\
Classic communication leakage from Alice (with filtering) & 8.6~cps (SNSPD1) and 8.5~cps (SNSPD2) \\
Classic communication leakage from Bob (with filtering)        & 6.7~cps (SNSPD1) and 6.2~cps (SNSPD2)  \\
Strong reference leakage   & 32~cps (SNSPD1) and 35~cps (SNSPD2) \\
Dark count (SNSPD1) & 4~cps  \\
Dark count (SNSPD2) & 6~cps  \\
\hline
\end{tabular}
\label{Tab:noise}
\end{table*}

\subsection*{The feedback details of the system}

In our experiment, besides the influence of various noises as mentioned above. Another question needs to be considered is that the vibration and temperature fluctuations caused by the ambient would indiscriminately change the polarization and fiber delay, which undoubtedly deteriorate the performance of single photon interference. Therefore, in Charlie, we optimize the noises of the system, and introduce a real-time feedback of the polarization and arrival time of the pulses from the two encoders.

First a 100 GHz Dense Wavelength Division Multiplexing (DWDM) at the central wavelength of 1550.12~nm is inserted to filter out the crosstalk noise from running fiber links of classical communication which are in the same bundle with the QKD link, and to filter out the inelastic scattering noise caused by strong reference pulses~\cite{chen2020sending}. In the following a circulator (CIR) is used to block the noise caused by the reflection of the four superconducting nanowire single-photon detectors (SNSPDs)~\cite{chen2020sending}, after that the pulses are polarized and split by the electric polarization controllers (EPCs) and polarization beam splitters (PBSs), the idle beam of the PBS is detected to carry out the arrival time and polarization feedback, another beam is sent into a polarization-maintaining beam splitter (BS) to accomplish the single photon interference. All the detections are recorded by a high speed multichannel time tagger.

\begin{figure*}[htb]
	\centering
	\resizebox{9cm}{!}
	{\includegraphics{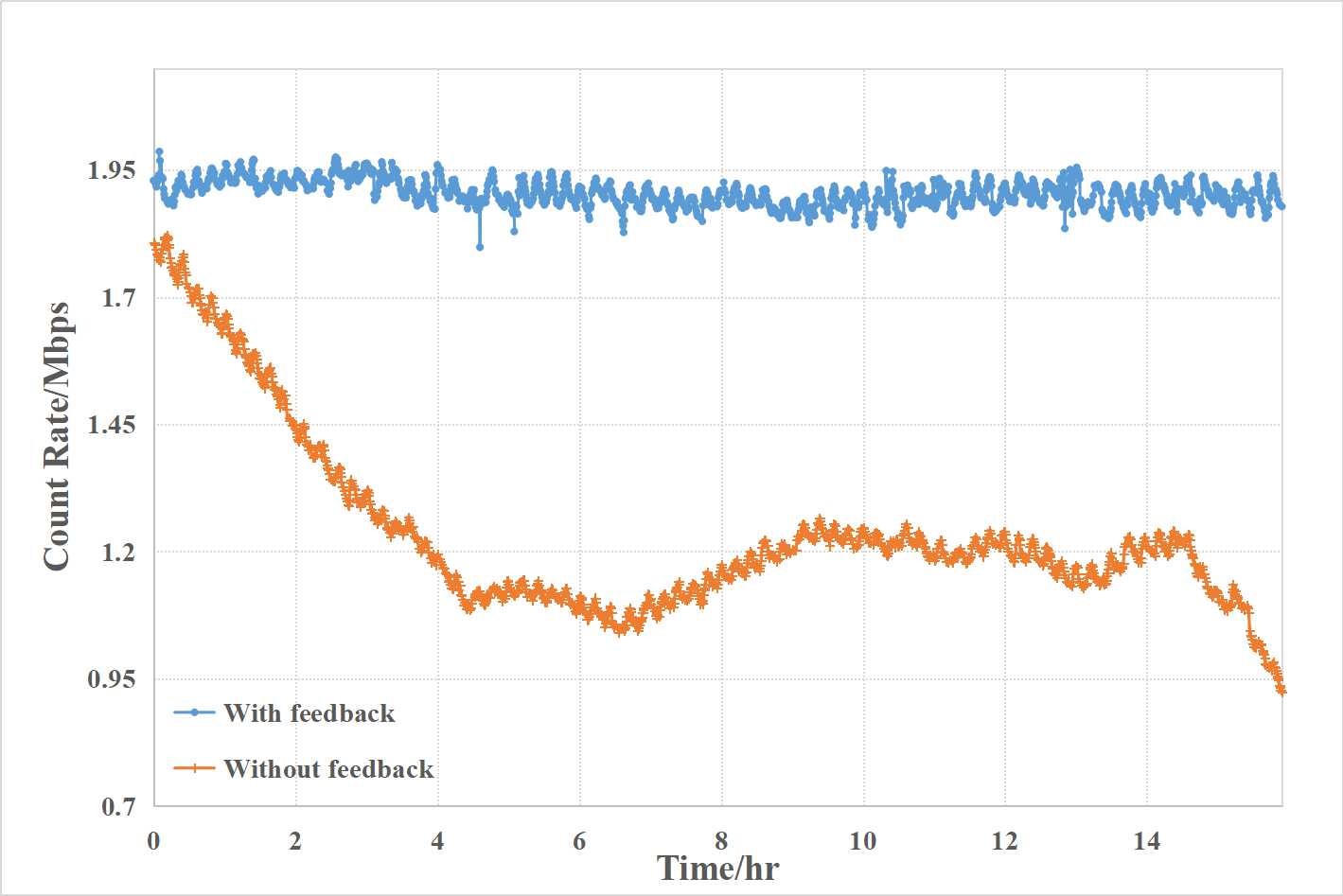}}
	\caption{\textbf{The variation of the polarization with and without feedback.}}
	\label{Fig:fluctuation}
\end{figure*}

Then, we measured the fluctuation of count rate caused by the variation of polarization. As shown in Fig.~\ref{Fig:fluctuation}, the blue curve shows the variation of the count rate without the feedback under field conditions, the intensity changes 30\% in 4 hours. Consequently, in order to improve the visibility of interference, we set a target count rate range of 150~kHz to 200~kHz of the idle beam of the PBS used for feedback, meanwhile adjust the EPC once a second to make the actual count rate meet it. The red curve shows the variation of the count rate with a feedback, which is almost stable. Additionally, the vibration and temperature fluctuations caused by the ambient would cause non-overlapping between the two pulses of interference. Also we measured the arrival time of the light pulses from one of the encoder (Bob) and found it is related to the temperature. As shown in Fig.~\ref{Fig:feedback}(a) in main text, the blue curve shows the variation of the arrival time, the red circle shows the temperature. Obviously, the arrival time is positively correlated with temperature, and the peak-to-peak value of the variation of the arrival time in a day is about 20 ns. To make the two pulses from Alice and Bob overlap each other, we set a target value to adjust the phase of the synchronization signals from Charlie's AWG once 20 seconds to make the actual arrival time meet the target. The orange curve shows the arrival time with a feedback, the according peak-to-peak value of variation of the arrival time in 6 hours is less than 80 ps.

\subsection*{Detailed Experimental Results}

Tab.~\ref{Tab:Characterization} shows the parameters for the loss of fiber and the transmittance of optical elements, as well as the SNSPD efficiencies. For the optical elements, we measured the optical transmittance including the polarization controller (PC), the dense-wavelength-division-multiplexer (DWDM), the circulator (CIR), the polarization beam splitter (PBS) and the polarization maintaining beam splitter (BS). Here, transmittances are given for each of the two inputs (A/B) and outputs (ch1/ch2), as appropriate. The SNSPD efficiency measurements included the PC efficiency.

\begin{table*}[htb]
\centering
  \caption{Experimental parameters of efficiencies of fiber and optical elements.}
\begin{tabular}{c|c}
\hline
Fiber Length    & 511 km  \\
$Loss_{FiberA}$	& 44.50dB     \\
$Loss_{FiberB}$	& 44.60dB   \\
\hline
PC-A & 99.1\%\\
PC-B & 99.1\%\\
\hline
DWDM-A & 93.1\%\\
DWDM-B & 92.0\%\\
\hline
CIR-A & 88.9\%\\
CIR-B & 86.9\%\\
\hline
PBS-A & 82.5\%\\
PBS-B & 77.5\%\\
\hline
BS-A-ch1 & 45.3\%\\
BS-A-ch2 & 44.4\%\\
BS-B-ch1 & 43.7\%\\
BS-B-ch2 & 44.0\%\\
\hline
SNSPD-ch1 & 77.0\%\\
SNSPD-ch2 & 91.0\%\\
\hline
\end{tabular}
\label{Tab:Characterization}
\end{table*}

Tab.~\ref{Tab:Result} summarizes the experimental results. It shows the total sending number of signal pulses ($N_{total}$) and the final key rate $R$ for the best possible accepted phase difference $Ds$ (in degrees). In our experimental implementation, we applied a digital window to select the signal in the middle of each pulse, where the interference is expected to be better and all the detections are filtered according to it, before the detections are announced by Charlie. The results of untagged bits in the $Z_1$-windows ($n_1$), phase-flip error rate $e_1^{ph}$ and the error rates in $Z$ basis before and after applying the AOPP method are also given respectively. ``QBER(X11)" and ``QBER(X22)" represent for the error rates of the decoy states ``11" and ``22".

In the following rows, we list the numbers of pulses Alice and Bob sent in different decoy states, labelled as ``Sent-ABCD", where ``A" (``B") is ``X" or ``Z" indicating the basis which Alice (Bob) has used; ``C" (``D") is ``0", ``1",  ``2" or  ``3", indicating the intensity which Alice (Bob) has chosen within ``vacuum", ``$\mu_1$", ``$\mu_2$" or ``$\mu_z$". As with the numbers of sent pulses, the numbers of detections are listed as ``Detected-ABCD". The numbers of valid detections in $Z$ basis before AOPP are listed as ``Detected-Valid-$Z$(Before AOPP)'' and the corresponding survived bits after AOPP are listed as ``Survived-bits(After Aopp)''. The total valid detections reported by Charlie is denoted as ``Detected-Valid-ch", where ``ch" can be ``Det1" or ``Det2" indicating the responsive detector of the recorded counts. The table also gives the numbers of detections falling within the accepted difference range $Ds$, listed as ``Detected-ABCD-Ds-Ch", where ``Ds" indicates that only the data within the accepted range $Ds$ are counted, ``Ch" indicates the detection channel. The numbers of correct detections are listed as ``Correct-ABCD-Ds-Ch", and used to calculate the $X$ basis error rate. ``Detected-ZZError" and ``Detected-ZZCorrect" mean the numbers of error and correct detections in $Z$ basis respectively.

\linespread{1.5}
\begin{table*}[htb]
\centering
  \caption{Experimental results.}
\begin{tabular}{c|c|c|c}
\hline
\hline
Fiber Length & 511 km &Fiber loss &89.1 dB\\
\hline
$Ds$		 & $10^\circ$  &$R$    & $3.45\times 10^{-8}$ \\
\hline
$N_{total}$	 & $1.679\times10^{12}$  &Sent-ZZ		 & 907428400000	\\
Sent-ZX00	 & 19393400000	 &Detected-ZX00	& 269	  	\\
Sent-ZX01	 & 204410800000	 &Detected-ZX01	& 294733	\\
Sent-ZX02	 & 21217400000	&Detected-ZX02	& 88940	\\
Sent-ZX30	 & 8707200000	&Detected-ZX30	& 49610			\\
Sent-XZ00	 & 21229000000	&Detected-XZ00	& 275	\\
Sent-XZ10	 & 198592000000	&Detected-XZ10	& 279328\\
Sent-XZ20	 & 23111200000	&Detected-XZ20	& 96684	\\
Sent-XZ03	 & 9517200000	&Detected-XZ03	& 51635			\\
Sent-XX00	 & 1394800000	&Detected-XX00	& 20	\\
Sent-XX01	 & 7057600000 &Detected-XX01	& 10150	\\
Sent-XX02	 & 0 &Detected-XX02	& 0\\
Sent-XX10	 & 2174200000 &Detected-XX10	& 3037\\
Sent-XX20	 & 1235800000 &Detected-XX20	& 5281\\
Sent-XX11	 & 79038400000 	&Detected-XX11	& 225307\\
Sent-XX22	 & 1517800000 &Detected-XX22	& 12519\\
\hline
$n_1$(Before AOPP)		 & 1255190 &$n_1$(After AOPP) & 219136 \\
$e_1^{ph}$(Before AOPP)  & 8.078\% &$e_1^{ph}$(After AOPP)   & 16.067\%\\
QBER$(Z-{Before})$ 	 & 24.393\%  &QBER$(Z-{After})$ 	 & 0.431\%  \\
Detected-Valid-$Z$(Before AOPP) &2631682&Survived-bits(After AOPP)&576130\\
QBER(X11)	 & 4.9\% &QBER(X22)	 & 5.8\% \\
\hline
Detected-Valid-Det1	& 2088506 &Detected-Valid-Det2	& 2898644 \\
Detected-XX11-Ds-Ch1	& 11162 &Detected-XX11-Ds-Ch2	& 15073\\
Correct-XX11-Ds-Ch1		& 10593 &Correct-XX11-Ds-Ch2		& 14353 \\
Detected-ZZError	& 641867 &Detected-ZZCorrect	& 1989815\\
\hline
\end{tabular}
\label{Tab:Result}
\end{table*}

Different accepted phase difference ranges $Ds$ lead to different detection counts and QBERs in $X$ basis. We list the QBERs of the $X$ basis when Alice and Bob send decoy states $\mu_1$ and $\mu_2$ with different phase difference range ($Ds$), which are listed in Tab.~\ref{Tab:ResultQBERXX11}. It also shows the different detection counts according to different $Ds$ and the optimized secure key rates that we extract by searching through ranges of these parameter values.

\begin{table*}[htb]
\centering
  \caption{QBERs and detections in $X$ basis and key rates with 511 km fiber.}
\begin{tabular}{c|cccccccc}
\hline
Results$\mid$Ds/2   & deg=2$^\circ$	& deg=8$^\circ$	& deg=10$^\circ$	& deg=12$^\circ$	& deg=15$^\circ$	& deg=30$^\circ$\\
\hline

QBER(X11)	& 4.8\% & 4.8\% & 4.9\% & 5.0\% & 5.2\% & 7.1\%  \\
QBER(X22)   & 3.4\% & 5.8\% & 5.8\% & 5.7\% & 5.2\% & 6.7\% \\
Detections of X11 & 6210 & 21432  & 26235 & 31562 & 39046 & 76935 \\
Detections of X22 & 348  & 1260 & 1511 & 1780 & 2206 & 4302 \\
Key Rates & $1.85\times 10^{-8}$ & $3.37\times 10^{-8}$ & $3.45\times 10^{-8}$ & $3.43\times 10^{-8}$ & $3.36\times 10^{-8}$ & $2.23\times 10^{-8}$  \\

\hline
\end{tabular}
\label{Tab:ResultQBERXX11}
\end{table*}

\end{appendix}
\clearpage

%%%%%%%%%%%%%%%%%%%%%%%%%%%%%%%%%%%%%%%%
% choose a style
%\bibliographystyle{ieeetr}
%\bibliographystyle{unsrt}
\bibliographystyle{apsrev}
\bibliography{BibSNSTFQKD}

\end{document}